\documentclass[twocolumn,showpacs,preprintnumbers,amsmath,amssymb]{revtex4}

\usepackage{CJK}

\usepackage{graphicx}

\usepackage{dcolumn}

\usepackage{bm}
\usepackage[normalem]{ulem}
\usepackage{color}

\usepackage{epstopdf}

\newcommand{\ii}{{\rm i}}

\newcommand{\bp}{\mathbf{p}}
\newcommand{\bq}{\mathbf{q}}
\newcommand{\bv}{\mathbf{v}}

\newcommand{\br}{\mathbf{r}}

\newcommand{\bff}{\mathbf{f}}

\newcommand{\bh}{\mathbf{h}}

\newcommand{\bk}{\mathbf{k}}

\newcommand{\sep}{ \ \ \ , \ \ \ }

\newcommand{\beq}{\begin{equation}}
\newcommand{\eeq}{\end{equation}}
\newcommand{\beqn}{\begin{eqnarray}}
\newcommand{\eeqn}{\end{eqnarray}}
\newcommand{\pp}{\partial}

\newcommand{\eq}{Eq.\ }

\newcommand{\eqs}{Eqs }
\newcommand{\fig}{Fig.\ }

\newcommand{\cO}{{\cal O}}

\newcommand{\cP}{{\cal P}}

\newcommand{\la}{\langle}

\newcommand{\tbq}{\tilde{\bq}}
\newcommand{\tbk}{\tilde{\bk}}
\newcommand{\tbp}{\tilde{\bp}}
\newcommand{\tbh}{\tilde{\bh}}
\newcommand{\ra}{\rangle}

\begin{document}

\begin{CJK*}{GBK}{}

\title{Critical Phenomenon of the Order-Disorder Transition in Incompressible Flocks}
\author{Leiming Chen}
\email{leimingyandongyu@gmail.com}
\address{College of Science, China University of Mining and Technology, Xuzhou Jiangsu, 221116, P. R. China}
\author{John Toner}
\email{jjt@uoregon.edu}
\affiliation{Department of Physics and Institute of Theoretical
Science, University of Oregon, Eugene, OR $97403^1$}
\affiliation{Max Planck Institute for the Physics of Complex Systems, N\"othnitzer Str. 38, 01187 Dresden, Germany}
\author{Chiu Fan Lee}
\email{c.lee@imperial.ac.uk}
\address{Department of Bioengineering, Imperial College London, South Kensington Campus, London SW7 2AZ, U.K.}
\date{\today}
\begin{abstract}
We study incompressible systems of motile particles with  alignment interactions. 
Unlike their  compressible  counterparts, in which the order-disorder (i.e., moving to static)  transition, tuned by either noise or number density, is discontinuous,
in incompressible systems this transition can be continuous, and belongs to a new universality class. We calculate 
the critical exponents
 to $\cO(\epsilon)$
in an  $\epsilon=4-d$ expansion, and derive
two exact scaling relations. 
This is the first analytic treatment of a phase transition in a new universality class  in an active system.
\end{abstract}
\maketitle
\end{CJK*}

%

Emergent properties of interacting  non-equilibrium systems are of widespread and fundamental interest.
One of the simplest, but most striking, of these is the self-organized phenomenon of ``flocking"- that is,  collective motion (CM) in large groups of motile organisms \cite{boids, Vicsek, TT1,TT2,TT3,TT4, dictyo, rappel1, actin, microtub}.
This phemenon is fascinating 
in part because its occurrence in two spatial dimensions requires the spontaneous breaking of a continuous symmetry,   which is forbidden in thermal equilibrium by the Mermin-Wagner theorem \cite{MW}.
It was initially hoped \cite{Vicsek} that the transition into this  novel state could be a continuous one belonging to  a new universality class. However, it was subsequently realized, from both simulations and  theoretical analysis \cite{band1,band2,band3,band4,band5,band6,band7} of the hydrodynamic equations \cite{TT1,TT2,TT3,TT4},  that as this putative continuous transition is approached from the ordered side, but before
it can  be reached, the homogeneous CM state becomes unstable to
modulation of the  density 
along the mean flock velocity.  
The transition from the  homogeneous CM state  (i.e., the ordered state) to the disordered state proceeds via two first order transitions: one from homogeneous to banded, the next from banded to disordered.

Since this instability   requires 
density variations, it 
can be eliminated by eliminating density fluctuations: that is, by making system incompressible. In this paper, we show that the order-disorder
transition is continuous in an incompressible system. We demonstrate this  by finding, in 
a dynamical renormalization group (DRG) analysis of the hydrodynamic equations for an incompressible flock in $d$ spatial dimensions, 
a stable fixed point that controls the transition. This calculation is done to order
$\cO(\epsilon)$ in an  $\epsilon=4-d$ expansion; to the same order,
we calculate
the critical exponents of the transition. We also obtain two 
scaling laws relating these critical exponents 
which are valid to all orders in $\epsilon$ (i.e., exact).

Our results are testable in both experiments and simulations. Three potential realizations are:

\noindent1) Systems  with strong repulsive short-ranged interactions between the active particles.  Incompressibility has, in fact, been assumed in, e.g.,   recent experimental studies on cell motility \cite{wensink_pnas12}.
In such systems, the compressibility will be non-zero,
but small.
Hence, our incompressible results will apply out to  very large length scales, or, equivalently, very close to the transition, but will ultimately crossover to the compressible behavior (i.e., 
 a small first order transition  driven 
 by the banding instability).

\noindent2) Systems with long-ranged repulsive interactions; here, true incompressibility is possible. Long ranged interactions are quite reasonable in certain contexts: birds, for example, can often see all the way across a flock \cite{turner}.


%
\noindent3) 
Motile colloidal systems in fluid-filled microfluidic channels.
The forces exerted by the active particles are, of course, tiny compared to what would be needed to compress the background fluid, so that fluid is effectively incompressible. Since the active particles drag the background fluid with them, their motion is effectively incompressible as well.  
Indeed, experiments 
\cite{bartolo} show these systems do not exhibit the banding instability \cite{band1, band2, band3, band4, band5, band6, band7} found in all compressible active systems. 
This also
suggests a numerical approach: 
simulating active particles moving through an incompressible fluid \cite{Stark}.


We formulate the most general hydrodynamic model for systems lacking both momentum conservation, and Galilean invariance,
consistent with the symmetries of rotation and translation invariance, and the assumption of incompressibility.
As the number density cannot fluctuate (by the assumption of incompressibility), 
the velocity field is  the only hydrodynamic variable in the problem, which becomes soft as the transition is approached. Since the velocity is small near the transition, we can expand  the equation of motion (EOM) in powers of the velocity. 
The  symmetry constraints of translation  and rotation invariance
force the EOM valid at long wavelengths and times to take the form: \cite{comm, TT1,TT2,TT3,TT4}
\begin{eqnarray}
\pp_t \bv +\lambda (\bv \cdot {\bf \nabla})\bv
 &=& -{\bf \nabla} \cP -(a+b |\bv|^2) \bv\nonumber\\
&&+\mu \nabla^2 \bv + \bff
\label{eq:original}.
\end{eqnarray}
where the pressure $\cP$ 
enforces the incompressibility condition ${\bf \nabla} \cdot \bv =0$,
$\bff$ is a ``white noise" with spatio-temporally Fourier transformed statistics:
\beq
\la f_m (\bk,\omega) f_n(\bk',\omega') \ra =2D P_{mn} (\bk)\delta(\bk+\bk')\delta(\omega +\omega')
\ ,
\eeq
and $ P_{mn} (\bk) \equiv \delta_{mn} -k_m k_n/k^2$ is the transverse projection operator.
%
This EOM  (\eq (\ref{eq:original})) reduces,
when $a=0=b$, to the classic model of a
fluid forced at zero wavenumber treated by \cite{FNS} (their ``model B''). With $\lambda=0$, it reduces to a simple, time-dependent Ginzburg-Landau (TDGL) \cite{Ma, Lubensky} dynamical model for an isotropic ferromagnet with long ranged dipolar interactions \cite{Aharony, comm2}.

Because our system lacks Galilean invariance, $\lambda$ need not (and in general will not) be one, and the terms $-(a+b |\bv|^2) \bv$ are allowed in the EOM.
The latter is crucial as it explains why there can be a polar ordered phase in an active system, which is not possible in a normal fluid.

\begin{figure}[b]
\begin{center}
\includegraphics[scale=.35]{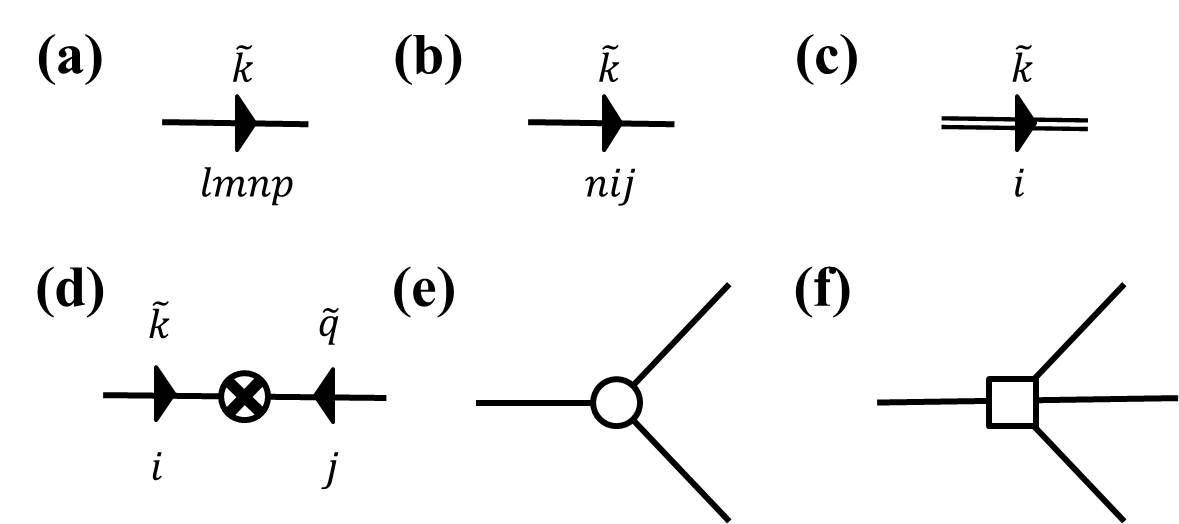}
\end{center}
\caption{Graphical representations: (a)$=Q_{lmnp}(\bk)G(\tilde{\bk})$; (b) $=P_{nij}(\bk)G(\tilde{\bk})$; (c) $=v_i(\tilde{\bk})$; (d) $=2DP_{ij}(\bk)\mid G(\tilde{\bk})\mid^2$; (e) $=-{\ii\over 2}\lambda$; (f) $=-{b\over 3}$.}
\label{fig:pic1}
\end{figure}

At the  mean field level, for $a<0,b>0$ the system is in the  ordered
phase with $|\bv|=\sqrt{-a/b}$,
and for $a>0,b>0$ it is in the disordered phase with $|\bv|={\bf 0}$.  
To go beyond this mean field description, we employ the DRG method \cite{FNS} near the order-disorder transition. To do so, we spatio-temporally Fourier transform  Eq. (\ref{eq:original}), and project orthogonal
to wavevector $\bk$;  obtaining
\beqn
v_l (\tilde{\bk}) = G(\tilde{\bk}) \Bigg[ f_l(\tilde{\bk})-\frac{\ii \lambda}{2} P_{lmn}(\bk) \int_{\tilde{\bq}} v_m(\tilde{\bq}) v_n(\tilde{\bk}-\tilde{\bq})\nonumber
\\
\label{RGmodel}
-\frac{b}{3} Q_{lmnp}(\bk) \int_{\tilde{\bq},\tilde{\bh}} v_m(\tilde{\bk}-\tilde{\bq}-\tilde{\bh}) v_n (\tilde{\bq})v_p (\tilde{\bh}) \Bigg]
\label{FTEOM}
\eeqn
where we have adopted the reduced notations $\tilde{\bk} \equiv (\bk,\omega)$ and $\int_{\tilde{\bq}} = \int_{\bq,\Omega} \equiv \int \frac{d^d q}{(2\pi)^d }\frac{d\Omega}{2\pi }$, and
we have defined
$P_{lmn}(\bk)\equiv P_{lm}(\bk) k_n +P_{ln}(\bk) k_m$,
$Q_{lmnp}(\bk) \equiv P_{lm}(\bk)\delta_{np}+P_{ln}(\bk)\delta_{mp}+P_{lp}(\bk)\delta_{mn}
$,
and
the ``propagator" $G(\tilde{\bk})\equiv(-\ii \omega +\mu k^2 +a)^{-1}$. 
Graphical representations of the various terms in \eq (\ref{RGmodel}) are shown in \fig \ref{fig:pic1}. 

\begin{figure}[b]
\begin{center}
\includegraphics[scale=.35]{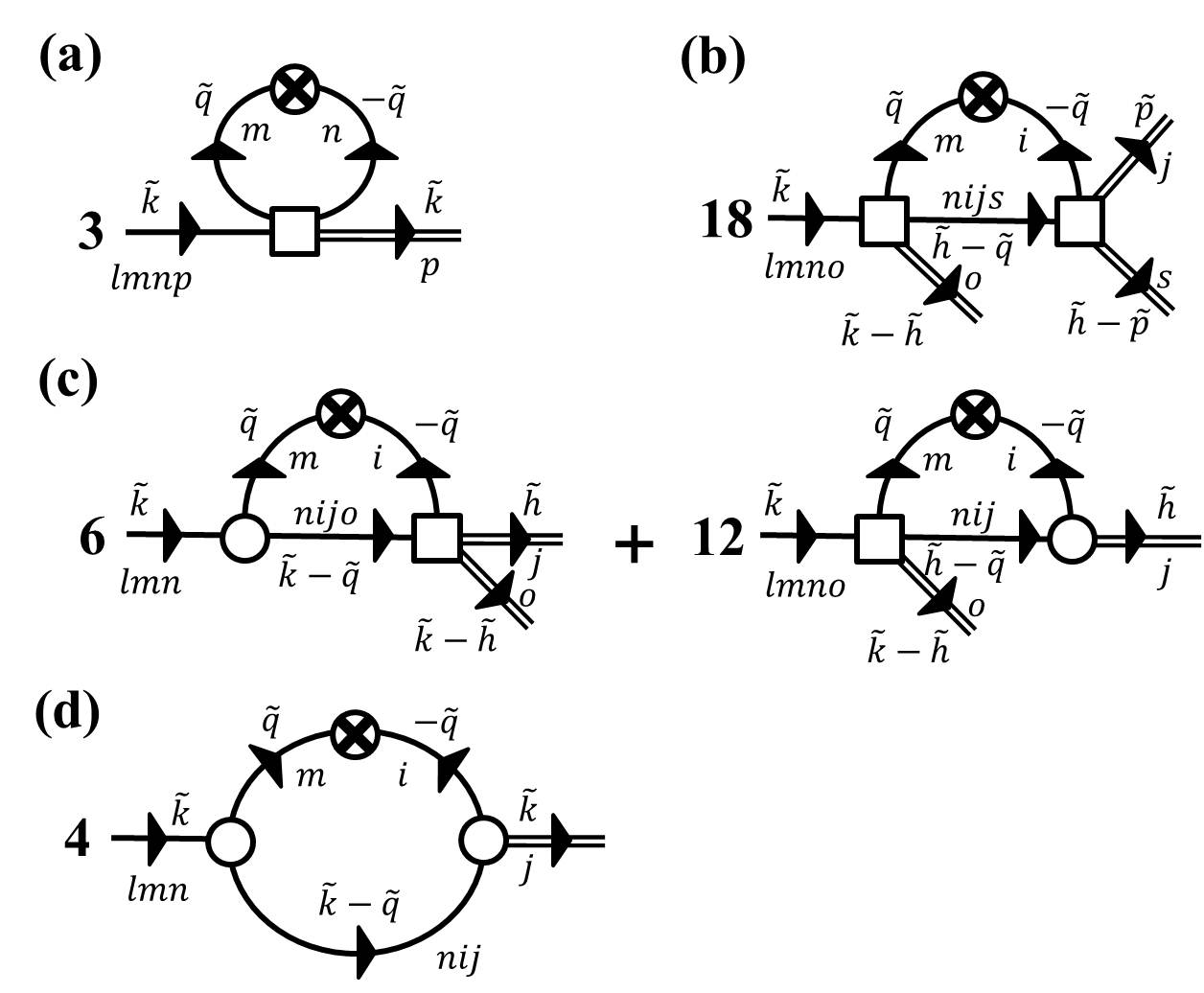}
\end{center}
\caption{Non-vanishing diagrams at the one-loop level. Diagrams (a) to (d)  contribute to  
$a$, $b$, $\lambda$ and $\mu$ respectively.}
\label{fig:pic2}
\end{figure}

%
We now perform the standard DRG procedure \cite{FNS}, averaging over short wavelength degrees of freedom, and  rescaling:
 ${\bf r}\to{\bf r}e^{\ell}$, $t\to te^{z\ell}$ and $\bv\to e^{\chi\ell}\bv$. Our procedure is identical to the calculation for model B in \cite{FNS}, except for a modified propagator, and some additional Feynmann graphs due to the extra $b |\bv|^2 \bv$ non-linearity in our problem. At the one loop level, the non-vanishing graphical contributions to the various coefficients in \eq (\ref{RGmodel}) are shown in \fig \ref{fig:pic2}. More details of the calculation 
 are given in the supplemental materials. We
obtain, in $d$ spatial dimensions,  the following RG flow equations of the coefficients to one loop order and to linear order in $\epsilon\equiv4-d$ \cite{supp}:
\begin{eqnarray}
 {da\over d\ell}&=&za-\frac{9g_2}{2}(a-\mu\Lambda^2),
 \label{alpha1}
 \\
  {db\over d\ell}&=&\left(2\chi+z-\frac{17g_2}{2}\right)b,
  \label{brr}\\
   {d\lambda\over d\ell}&=&\left(\chi-1+z-\frac{5g_2}{3}\right)\lambda,
   \label{lambdarr}\\
 {d\mu \over d\ell}&=&\left(-2+z+\frac{g_1}{4}\right)\mu,
 \label{murr}\\
 {dD\over d\ell}&=&\left(-2\chi+z-d\right)D
 \label{Drr} \ .
\end{eqnarray}
where we've defined dimensionless couplings:
\begin{eqnarray}
g_1\equiv {S_d D\lambda^2 \over\left (2\pi \right)^d \mu^3}\Lambda^{-\epsilon},\,\,\,
g_2\equiv {S_d Db\over\left (2\pi \right)^d\mu^2}\Lambda^{-\epsilon}\,,
\label{eq:g1g2}
\end{eqnarray}
and where  $S_d \equiv 2\pi^{d/2} /\Gamma (d/2)$ is the surface area of a unit sphere in $d$ dimensions, $\epsilon \equiv 4-d$,  and $\Lambda$ is the ultraviolet wavevector cutoff. Since our interest is in the transition, we have, in the last four recursion relations (\ref{brr}-\ref{Drr}),  set $a=0$, and have worked to linear order in $a$ in (\ref{alpha1}). It is straightforward to verify that higher order terms in $a$ affect none of our results up to and including linear order in $\epsilon=4-d$.


From these RG flow equations, we can derive two closed flow equations for $g_{1,2}$ for arbitrary $\chi$ and $z$:
\begin{eqnarray}
 {dg_1\over d\ell}&=&\epsilon g_1-\frac{3}{4}g_1^2-\frac{10}{3}g_1g_2 \ ,\label{g1}\\
 {dg_2\over d\ell}&=&\epsilon g_2-\frac{1}{2}g_1g_2-\frac{17}{2}g_2^2\ .\label{g2}
\end{eqnarray}
Although not necessary, it is convenient to make a special choice of $z$ and $\chi$ such that $\mu$ and $D$ are kept fixed at their bare values (i.e, $\mu_0$ and $D_0$, respectively). We will  hereafter adopt this choice of $z$ and $\chi$, which is
\beq
z=2-g_1/4+\cO(\epsilon^2)\,\,,\chi=\frac{z-d}{2}+\cO(\epsilon^2)\,.
\label{choice}
\eeq
We will also hereafter use  the subscript $0$ to denote the bare (i.e., unrenormalized) values of the parameters.

\begin{figure}[b]
\begin{center}
\includegraphics[scale=.45]{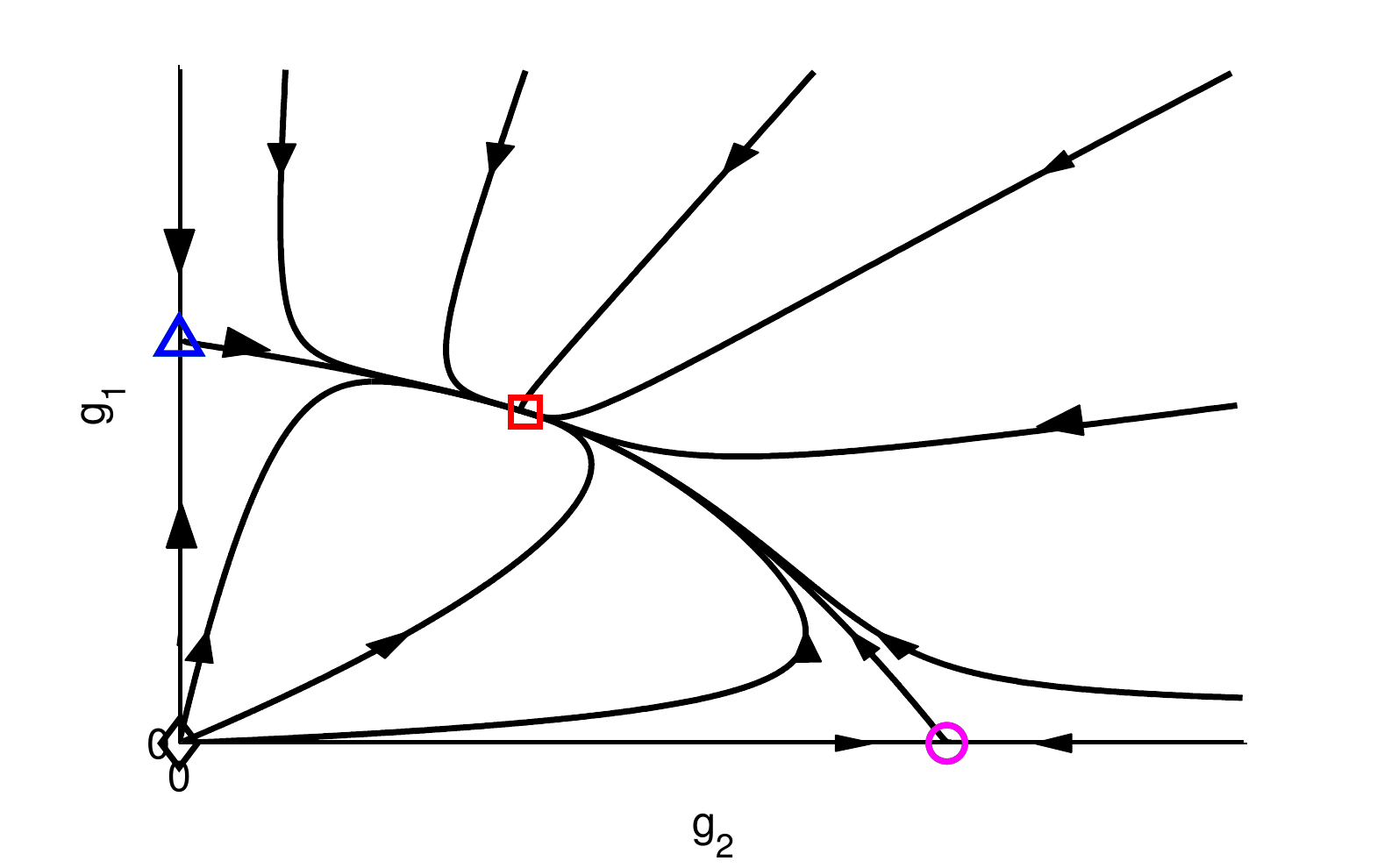}
\end{center}
\caption{RG flows on the critical surface. Besides the unstable Gaussian fixed point (black diamond) and the stable fixed described in \eqs (\ref{stableFP}) (red square), there are two unstable fixed points: one at $g^*_1=0$, $g^*_2=\frac{2\epsilon}{17}$, which is the fixed point of an isotropic ferromagnet with long-ranged dipolar interactions \cite{Aharony} (purple circle), and one at $g^*_2=0$,  $g^*_1=\frac{4\epsilon}{3}$, which is the fixed point of
a fluid forced at zero wavevector (Model B of \cite{FNS}) (blue triangle). }
\label{fig:rgflows}
\end{figure}

Eq. (\ref{alpha1}) now becomes
\beqn
{da \over d\ell}&=&\left(2-\frac{g_1}{4} -\frac{9}{2}g_2\right)a+\frac{9}{2}g_2\mu\Lambda^2 \ .
\label{alpha2}
\eeqn
Eqs.~(\ref{g1},\ref{g2},\ref{alpha2}) have a non-Gaussian  fixed point  in $d<4$:
\begin{eqnarray}
g_1^*&=&
               {124\over 113}\epsilon+\cO(\epsilon^2) \sep
g_2^*
={6\over 113}\epsilon+\cO(\epsilon^2) \nonumber\\
 a^*
&=&\left[-{27\over 226}\epsilon+\cO(\epsilon^2)\right]\mu\Lambda^2,
\label{stableFP}
\end{eqnarray}
which can be shown by analyzing the three recursion relations
to be a stable  attractor of all points on a  two-dimensional surface  (the ``critical surface'' ) in the three-dimensional parameter space $(g_1,g_2, a)$, but to be unstable with respect to displacements off this critical surface. The flows on the critical surface are illustrated in \fig \ref{fig:rgflows}. This is exactly the topology of renormalization group flows that corresponds to a continuous phase transition with universal exponents controlled by the fixed point that's stable within the critical surface. Hence, we conclude that  the order-disorder
{\it is} generically  continuous in incompressible active fluids.  

The exponential runaway
from the critical surface in the
unstable direction near the stable fixed point (\ref{stableFP})
grows like $e^{y_a\ell}$, with the exponent
\begin{eqnarray}
y_a
=2-{58\over 113}\epsilon+\cO(\epsilon^2).
\label{thermalEV}
\end{eqnarray}
This eigenvalue determines the critical exponent $\nu$ governing the critical behavior of the velocity correlation length $\xi$. In addition, the smallest (in magnitude) of the two negative eigenvalues gives the ``correction to scaling exponent" $y_2$ \cite{Ma}; 
we find 
$y_2
=-\frac{31}{113}\epsilon+\cO(\epsilon^2). $



A useful experimental probe of the transition is the velocity correlation function
$\langle \bv(\br+{\bf R},t+T)\cdot\bv({\bf R},T)\rangle\equiv C(\br,t)$, which depends on bare parameters  $b_0$, $\mu_0$, $D_0$,  and $\lambda_0$ and, most importantly, the proximity to the phase transition $\delta a_0\equiv a_0-a_0^c$, where $a_0^c$ is the value of $a_0$ at the transition. 
The RG 
connects the original $C(\br,t)$ to that of the rescaled system:
\begin{eqnarray}
&&C(\br, t; \delta a_0,b_0,\lambda_0)\nonumber\\
&&=e^{2\chi\ell}C \Big(\br e^{-\ell}, t e^{-z\ell}; \delta a(\ell),b(\ell),\lambda(\ell)\Big),
\label{RGconnection}
\end{eqnarray}
where we have not displayed $\mu_0$ and $D_0$, since they are kept fixed in the RG. By choosing $\ell=\ln(\Lambda r)$, and using $\delta a(\ell)\approx\delta a_0e^{y_a\ell}$, we can obtain from this a scaling form for large $r$:
\beq
C(\br, t)
=r^{2-d-\eta}Y_\pm\left( {r\over\xi}, {t\over r^z}\right) \ .
 \label{Correlation1}
\eeq
In particular, the equal time correlation function scales as $r^{2-d-\eta}$. In \eq \ref{Correlation1},
the exponent  $\eta$
is given by
\beq
 \eta=2-d-2\chi
=2-z+\cO(\epsilon^2)
 ={31\over 113}\epsilon+\cO(\epsilon^2)\ ,\label{eta}
 \eeq
the scaling functions $Y_\pm$ by
\beq
Y_\pm(x,y)=C( \Lambda^{-1},y; \pm x^{y_a}, b^*, \lambda^*)\,,
\label{scalingfn}
\eeq
and the diverging correlation length $\xi$ by $\xi\equiv\Lambda^{-1}|\delta a_0|^{-\nu}$, where the correlation length exponent
\begin{eqnarray}
\nu={1\over y_a}={1\over 2}+{29\over 226}\epsilon+\cO(\epsilon^2).
\end{eqnarray}
In our expression for the $\epsilon$ expansions for $\eta$, we have replaced $\chi$ and $z$ by their values at the fixed point (\ref{stableFP}), which is valid given that $r$ is large and the system is sufficiently close to the transition. We do so consistently when calculating other exponents as well.
The first line of equation (\ref{eta}) is exact (i.e., independent of the $\epsilon$-expansion), as it is simply the definition of $\eta$.

The order-disorder transition can be driven by tuning any one of many 
microscopic control parameters (e.g., density or noise strength).
Whatever control parameter $s$ is  tuned, we expect $a_0-a_0^c\propto \left(s-s_c\right)$ by   analyticity near $s_c$,
where $s_c$ is the values of  the control parameter
 $s$ at the transition.
 As a result, the velocity correlation length $\xi$ just defined diverges as $
 \xi\propto |s-s_c|^{-\nu}$
as {\it any} control parameter $s$ is tuned.

The scaling functions $Y_\pm$ in equation (\ref{Correlation1}) are different on the disordered ($+$) and ordered ($-$) sides of the transition, because the system is in different phases in the two cases. On the disordered side , we expect $Y_+(x,y)$ to decay exponentially with both $x$ and $y$, while on the ordered side, $Y_-(x,y)$ has a more complicated scaling behavior that we'll discuss elsewhere \cite{future}.

Now we calculate the magnitude of the order parameter in the ordered state near the critical point. The RG connects the average velocity of the original system and that of the rescaled system
with the relation
\beq
\langle\bv\rangle (\delta a_0, b_0, \lambda_0)=e^{\chi\ell} \langle\bv\rangle \left(  \delta a_0e^{y_a\ell}, b(\ell), \lambda(\ell)\right)\ .
\label{orderRG1}
\eeq
We choose $\ell$ such that $\delta a_0e^{y_a\ell}$ is of order 1.  Therefore, $\ell$ is large since $\delta a_0$ is small near the critical point, and hence, both $b(\ell)$ and $\lambda(\ell)$ flow to their nonzero fixed values. Then all the singular dependence on  $(s_c-s)$ on the RHS
of the equality (\ref{orderRG1}) is included in the exponential. This implies $ |\langle\bv\rangle|\sim |s_c-s |^{\beta}$
with
\beq
\beta=-\nu\chi={1\over 2}-{6\over 113}\epsilon +\cO(\epsilon^2)\ .
\label{beta}
\eeq

The first equality in this expression, which is {\it exact}, can be rewritten in terms of $\eta$ using the definition of $\eta$ embodied in the first equality of (\ref{eta}), giving the {\it exact}  hyperscaling relation
\begin{eqnarray}
\beta&=&{\nu\over 2}(d-2+\eta)\ .
\label{exact1}
\end{eqnarray}
In this respect, our system is similar to  equilibrium systems, in which (\ref{exact1}) also holds \cite{Lubensky}.

We study next the linear response of the system to  a weak external field $\mathbf H$; that is, simply adding a small constant vector  $\mathbf H$ to the RHS of (\ref{eq:original}). In this case, the RG leads to the scaling relation
\beq
\langle\bv\rangle (\delta a_0, b_0, \lambda_0, H)=e^{\chi\ell} \langle\bv\rangle \left(  \delta a_0e^{y_a\ell}, b(\ell), \lambda(\ell), He^{y_H\ell}\right)\ ,
\label{orderRG2}
\eeq
where $H\equiv|\mathbf H|$ and $y_H$ is the RG eigenvalue of the external field $H$ at the fixed point (\ref{stableFP}). As there are no one loop graphical corrections to the external field, we can obtain $y_H$ to  $\cO(\epsilon)$ by simple power counting, which gives
\beq
y_H=z-\chi+\cO(\epsilon^2)\,.
\label{y_H}
\eeq
Again choosing $\ell$ such that  $\delta a_0e^{y_a\ell}$ is of order 1, we obtain
\beq
\langle\bv\rangle (\delta a_0, b_0, \lambda_0, H)=\left(\Lambda\xi\right)^\chi \langle\bv\rangle \left(  1, b^*, \lambda^*, H\left(\Lambda\xi\right)^{y_H}\right)\,,
\label{gammaRG1}
\eeq
where $b^*$ and $\lambda^*$ are the nonzero fixed values of $b(\ell)$ and $\lambda(\ell)$, respectively. Since the expectation value on the right hand side is evaluated in a system far from its critical region (since $\delta a=1$), we expect linear response to the external field on that side with an order one susceptibility. Hence,
\beq
\langle\bv\rangle (\delta a_0, b_0, \lambda_0, H)\sim\left(\Lambda\xi\right)^\chi  H\left(\Lambda\xi\right)^{y_H\ell}\propto\xi^{\chi+y_H}H\,,
\label{gammaRG2}
\eeq
which implies a linear susceptibility $\chi_H$ which diverges as $|s-s_c|^{-\gamma}$ with
\begin{eqnarray}
\gamma=\nu (\chi+y_H)=\nu z+\cO(\epsilon^2)=1+{27\over 226}\epsilon+\cO(\epsilon^2)\, .\label{gamma}
\end{eqnarray}
Note that Eqs.~(\ref{eta},\ref{gamma}) seem to suggest that $\eta$ and $\gamma$ satisfy Fisher's scaling law $\gamma=(2-\eta)\nu$.
However, since our system is out of equilibrium and thus the fluctuation dissipation theorem is not expected to hold,
we do {\it not} expect Fisher's scaling law to hold; indeed, the $\cO(\epsilon^2)$ terms probably violate it. The first line of equation (\ref{gamma}) {\it is} exact, however, and can be used to derive another scaling law, as we'll now show.
Turning on a small field right at the transition, we can again relate then the average velocities of the original and the rescaled systems using \eq (\ref{orderRG2}).
However,$\delta a(\ell)$ now flows to $0$  for large $\ell$ since the system is right at the critical point.
Therefore, by choosing $\ell=\ln\left(1/H\right)/y_H$, we obtain the $H$-dependence of $\bv$:
\beq
\langle\bv\rangle ( \delta a_0, b_0,\lambda_0,H)
=H^{-{\chi\over y_H}} \langle\bv\rangle \left(0, b^*, \lambda^*,1\right)\ .\ \ \ \label{}
\eeq
This implies
$\delta=-{y_H\over \chi}$.
Combining this with the exact first equalities in \eqs (\ref{beta},\ref{gamma}),  
we obtain  Widom's scaling relation
\begin{eqnarray}
\gamma=\beta(\delta-1) ,
\label{Widom}
\end{eqnarray}
which is {\it exact}. Plugging the
$\epsilon$-expansions of $\gamma$ and $\beta$
 into this relation, we find
\begin{eqnarray}
\delta=3+{51\over 113}\epsilon+\cO(\epsilon^2)\, .
\label{delta2}
\end{eqnarray}

In summary, we have studied the order-disorder transition in incompressible flocks using a dynamical $\epsilon=4-d$ expansion. This is the first study of the static to moving  phase transition in active matter
to go beyond mean-field theory, and include the effects of fluctuations on the transition.
We find a stable non-Gaussian fixed point, which implies a continuous transition, whose critical exponents were calculated to $\cO(\epsilon)$.
This fixed point is new, and so, therefore, is the universality class of this transition.
In addition, we found that the critical exponents obey two {\it exact} scaling relations which are the same as those in equilibrium ferromagnetic transitions, despite the fact that our system is fundamentally nonequilibrium. We also presented predictions for the scaling behavior of the velocity correlation functions.
Future theoretical work \cite{future} on this problem will include working out in quantitative detail the cutoff of the continuous transition by the banding instability in systems with a small, but non-zero, compressibility.

J.T. thanks Nicholas Guttenberg for explaining how to simulate systems with long ranged interactions; and the Max Planck Institute for the Physics of Complex Systems, Dresden; the Aspen Center for Physics, Aspen, Colorado; the Isaac Newton Institute, Cambridge, U.K., and the Kavli Institute for Theoretical Physics, Santa Barbara, California; for their hospitality while this work was underway.  He also thanks the  US NSF for support by
awards \# EF-1137815 and 1006171; and the Simons Foundation for support by award \#225579. L.C. acknowledges support by the National Science Foundation of China (under Grant No. 11474354).

\newpage
\section*{Supplmental Material}

The various constituent elements of Feynman diagrams are illustrated in Fig. 1 in the main text. 
The one loop graphical corrections to various coefficients in the model Eq. (1) in the main text
are illustrated in {Fig. 2
in the main text. Since we are only interested in the RG flows near the critical point, we have evaluated all of these graphs at $a=0$, except for the graph  (a) for $a$, the $a$-dependence of which we need to determine the ``thermal" eigenvalue $y_a$. The corrections from graphs  (a), (b), (c), and (d) are given respectively by:  
\begin{eqnarray}
 \delta a&=&{(d-1)(d+2)\over d}{S_d\over\left(2\pi\right)^d}{Db\over \mu\Lambda^2+a}d\ell,
 \label{delasum}
 \\
 \delta b&=&-\left[d+1+{6(d^2-2)\over d(d+2)}\right]{S_d\over\left(2\pi\right)^d}{b^2D\over \mu^2}\Lambda^{d-4}d\ell,\label{delb}
\\
\delta\lambda&=&-{2(d^2-2)\over d(d+2)}{S_d\over\left(2\pi\right)^d}{Db\lambda\over \mu^2}\Lambda^{d-4}d\ell\nonumber
\label{dellambdaIII}
\\
&&- {(d-2)\over d}{S_d\over\left(2\pi\right)^d}{Db\lambda\over \mu^2}\Lambda^{d-4}d\ell.\label{dellambdaIV}
 \\
\delta \mu&=&{(d-2)\over 2d}{S_d\over\left(2\pi\right)^d}{D\lambda^2\over \mu^2}\Lambda^{d-4}d\ell,
\label{delmu}
\end{eqnarray}

Combining these corrections (including the zero correction to $D$ ) with the rescalings described in the main text, {and setting dimension $d=4$ in all of these expressions (which is sufficient to obtain results to first order in $\epsilon=4-d$),  leads to the recursion relations  (4--8) in the main text,
(although in (4)
we have in addition expanded to linear order in $a$, which is sufficient to determine the exponents to $\cO(\epsilon)$).


\section{\label{Sec: graphs}Evaluation of the Feynman diagrams}


\subsection{Graph (a)}


This graph represents an additional
contribution $\delta \left(\partial_tv_l\right)$ to $ \partial_tv_l$ given by:
\beqn
\delta\left( \partial_tv_l\right)&=&-b Q_{lmnp}(\bk) v_p(\tbk)\int_{\bq, \Omega} \frac{2DP_{mn}(\bq)}{(\mu q^2+ a)^2+\Omega^2}\nonumber
\\ &=&-b DQ_{lmnp}(\bk)v_p(\tbk)\int_{\bq} \frac{P_{mn}(\bq)}{\mu q^2+ a}\nonumber\\
&=&-b DQ_{lmnp}(\bk)v_p(\tbk)\int_{\bq} \frac{\delta_{mn}(1-{1\over d})}{\mu q^2+ a}\nonumber\\
&=&-(d+2)b Dv_l(\tbk)\left(1-\frac{1}{d}\right)\int_{\bq} \frac{1}{\mu q^2+ a}
\nonumber\\
&=& -(d+2)b Dv_l(\tbk)\left(1-\frac{1}{d}\right){S_d\over\left(2\pi\right)^d} \frac{\Lambda^dd\ell}{\mu \Lambda^2+ a}\,,
\label{dela}
\eeqn
where $P_{mn}(\bq)=\delta_{mn}-q_mq_n/q^2$ is the transverse projection operator, and $Q_{lmnp}$ is defined in the main text after Eq. (3).
In going from the second to the third line above, we have used the well-known identity $\langle P_{mn}(\bq)\rangle_{_{\hat{\bf q}}}=\left(1-{1\over d}\right)\delta_{mn}$, where $\langle \rangle_{_{\hat{\bf q}}}$
denotes the average over directions ${\hat{\bf q}}$ of $\bq$ for fixed $|\bq|$.
This identity is derived in part (\ref{Sec: <}) of these Supplemental Materials.
Clearly this correction  to $\partial_t v_l(\bk, \omega)$ is 
exactly what one would obtain by adding to the parameter $a$ (hiding in the ``propagator'') in Eq. (3) in the main text
a correction $\delta a$ given by 
the coefficient of $v_l$ in (\ref{dela}); i.e., 
Eq. (\ref{delasum}).


\subsection{Graph (b)}

This graph represents an additional
contribution $\delta\left(\partial_tv_l\right)$ to $ \partial_tv_l$ given by:
\begin{widetext}
\beqn
\delta \left(\partial_tv_l\right)=18\left(\frac{b}{3}\right)^2 2DQ_{lmno}(\bk)\int_{\tbp,\tbh} v_j(\tbp)v_o(\tbk-\tbh)v_s(\tbh-\tbp)
\int_{\bq, \Omega} \frac{P_{mi}(\bq)Q_{nijs}(\bk-\bq)}{(\mu^2q^4+\Omega^2)[\mu |\bh-\bq|^2-\ii(\omega_h-\Omega)]}\ .
\label{dvVstart}
\eeqn
\end{widetext}
where the combinatoric prefactor of $18$ arises because there are $3$ ways to pick the leg with index $o$ on the left. Then, once this choice has been made, there are $2$ ways to pick the leg with index $m$ on the left, and $3$ ways to pick the one with index $i$ on the right.

To lowest order in the external momenta ($\bk$, $\bp$ and $\bh$) and frequencies $\omega$, $\omega_p$, and $\omega_h$, we can set all of these external momenta and frequencies equal to zero in the integrand of the integral over $\tbq$ in Eq. (\ref{dvVstart}). This proves, as we shall see, to make this graph into a renormalization
of the cubic non-linearity $b$. Making this simplification, Eq. (\ref{dvVstart}) becomes, after using
\beqn
\frac{1}{\mu q^2+\ii\Omega}=\frac{\mu q^2-\ii\Omega}{(\mu^2q^4+\Omega^2)}\,,
\label{propsimp}
\eeqn
and dropping an integral of an odd function of $\Omega$,
\begin{widetext}
\beqn
\delta \left(\partial_tv_l\right)=4b^2 DQ_{lmno}(\bk)\int_{\tbp,\tbh} v_j(\tbp)v_o(\tbk-\tbh)v_s(\tbh-\tbp)
\int_{\bq, \Omega} \frac{\mu q^2P_{mi}(\bq)Q_{nijs}(\bq)}{(\mu^2q^4+\Omega^2)^2}\ .
\label{dvV.2}
\eeqn
\end{widetext}
Using the definition 
of $Q_{nijs}(\bq)$, and contracting indices, we obtain
\begin{widetext}
\beqn
P_{mi}(\bq)Q_{nijs}(\bq)=P_{mn}(\bq)\delta_{js}+P_{ms}(\bq)P_{jn}(\bq)+P_{jm}(\bq)P_{ns}(\bq)\ .
\label{tensorV}
\eeqn
\end{widetext}
As we've done repeatedly throughout these Supplemental Materials, we'll replace this tensor with its average over all directions of $\hat{\bq}$. This is easily obtained from the known direction  averages   (\ref{P2angle}) and (\ref{P1}), and gives, after a little algebra,
\begin{widetext}
\beqn
\langle P_{mi}(\bq)Q_{nijs}(\bq)\rangle_{_{\hat{\bf q}}}=\frac{d+1}{d+2}\delta_{mn}\delta_{js}+\frac{d^2-2}{d(d+2)}\left(\delta_{ms}\delta_{jn}+\delta_{jm}\delta_{ns}\right)\ .
\label{angleV}
\eeqn
\end{widetext}
Inserting this into our earlier expression for $\delta\left( \partial_tv_l\right)$ and performing a few tensor index contractions gives
\begin{widetext}
\beqn
\delta\left( \partial_tv_l\right)&=&4b^2 D\left(\frac{d+1}{d+2}\delta_{js}Q_{lmmo}(\bk)+\frac{d^2-2}{d(d+2)}\left(Q_{lsjo}(\bk)+Q_{ljso}(\bk)\right)\right)\nonumber\\
&&\times\int_{\tbp,\tbh} v_j(\tbp)v_o(\tbk-\tbh)v_s(\tbh-\tbp)
\int_{\bq, \Omega} \frac{\mu q^2}{(\mu^2q^4+\Omega^2)^2}\,.
\label{dvV.3}
\eeqn
\end{widetext}
The trace in this expression can be evaluated as
\beqn
\nonumber
Q_{lmmo}(\bk)&=&P_{lm}(\bk)\delta_{mo}+P_{lm}(\bk)\delta_{mo}+P_{lo}(\bk)\delta_{mm}
\\
&=&(d+2)P_{lo}(\bk)\,,
\label{traceV}
\eeqn
and the integral over frequency $\Omega$ and $\bq$ is readily evaluated, and is given by
\beqn
\int_{\bq, \Omega} \frac{\mu q^2}{(\mu^2q^4+\Omega^2)^2}={1\over 4\mu^2}{S_d\over\left(2\pi\right)^d}\Lambda^{d-4}d\ell\,.
\label{intV}
\eeqn
Putting (\ref{traceV}) and (\ref{intV}) into (\ref{dvV.3}), and taking advantage of the complete symmetry of $\int_{\tbp,\tbh} v_j(\tbp)v_o(\tbk-\tbh)v_s(\tbh-\tbp)$ under interchanges of the indices $j$, $o$, and $s$ to symmetrize  the tensor prefactor gives
\begin{widetext}
\beqn
\delta\left( \partial_tv_l\right)=\frac{b^2 D}{3\mu^2}\left(d+1+6\frac{(d^2-2)}{d(d+2)}\right){S_d\over\left(2\pi\right)^d}\Lambda^{d-4}d\ell Q_{ljso}(\bk)\int_{\tbp,\tbh} v_j(\tbp)v_o(\tbk-\tbh)v_s(\tbh-\tbp)
\,.
\label{dvV.4}
\eeqn
\end{widetext}
This is readily recognized as a contribution to the $-\frac{b}{3}$ term in Eq. (3) in the main text; hence, the correction to $b$ is given by (\ref{delb}).

\subsection{The First Graph in (c)}

This graph represents an additional
contribution $\delta \left(\partial_tv_l\right)$ to $ \partial_tv_l$ given by:
\begin{widetext}
\beqn
\delta \left(\partial_tv_l\right)=6\left(-\frac{i\lambda}{2}\right)\left(-\frac{b}{3}\right)P_{lmn}(\bk)\int_{\tbh}v_j(\tbh)v_s(\tbk-\tbh)\int_{\bq, \Omega} Q_{nijs}(\bk-\bq)\frac{2DP_{mi}(\bq)}{(\mu^2q^4+\Omega^2)[\mu |\bk-\bq|^2-\ii(\omega-\Omega)]}\,,
\label{IIIstart}
\eeqn
\end{widetext}
where $P_{\ell mn}(\bk)$ is defined in the main text after Eq. (3), and the combinatoric prefactor of $6$ arises because there are $2$ ways to pick the leg with index $m$ on the left, and $3$ ways to pick the one with index $i$ on the right.

The piece of  $\delta \left(\partial_tv_l\right)$  linear in the external momentum $\bk$ is immediately recognized as a contribution to the $-{i\over 2}\lambda$ term in Eq. (3) in the main text. Since there is already an implicit factor of $\bk$ in the $P_{lmn}(\bk)$ in this expression, we can evaluate this graph to linear order in $\bk$ by setting both $\bk$ and the external frequency $\omega$  to zero in the integrand of the integral over $\tbq$. Doing so, and in addition using
(\ref{propsimp})
gives, after dropping an integral of an odd function of $\Omega$ that vanishes,
\begin{widetext}
\beqn
\delta\left( \partial_tv_l\right)=2i\lambda bDP_{lmn}(\bk)\int_{\tbh}v_j(\tbh)v_s(\tbk-\tbh)\int_{\bq, \Omega} \frac{\mu q^2P_{mi}(\bq)Q_{nijs}(\bq)}{(\mu^2q^4+\Omega^2)^2}\,,
\label{III.2}
\eeqn
\end{widetext}
where we've used the fact that $Q_{nijs}(\bq)$ is an even function of $\bq$. Using the definition of $Q_{nijs}(\bq)$, and performing the tensor index contractions, we can simplify the numerator of the integrand as follows:
\beqn
P_{mi}(\bq)Q_{nijs}(\bq)=P_{mn}\delta_{js}+P_{jn}P_{ms}+P_{ns}P_{mj}\,,
\label{QP}
\eeqn
where we've also used the fact that, e.g., $P_{mi}(\bq)P_{in}(\bq)=P_{mn}(\bq)$ (which is a consequence of the definition of $P_{mi}(\bq)$ as a projection operator). 
Using this result (\ref{QP}), and taking the angle average of the numerator of (\ref{III.2}) (which is the only factor in the integral that depends on the direction of $\bq$) gives,
\begin{widetext}
\beqn
\langle P_{mi}(\bq)Q_{nijs}(\bq)\rangle_{_{\hat{\bf q}}}=\left(\frac{d+1}{d+2}\right)\delta_{mn}\delta_{js}+\left(\frac{d^2-2}{d(d+2)}\right)\left(\delta_{jn}\delta_{ms}+\delta_{ns}\delta_{mj}\right)\,.
\label{QPave}
\eeqn
\end{widetext}
In deriving this expression, we've made liberal use 
of the angle averages (\ref{P1}) and (\ref{P2angle}).

The first term on the right hand side of (\ref{QPave}) contributes nothing, since it contracts two of the indices on the prefactor $P_{lmn}(\bk)$ together, which gives zero, as can be seen from the definition of $P_{lmn}(\bk)$:
\beqn
P_{lmm}(\bk)=P_{lm}(\bk)k_m+P_{lm}(\bk)k_m=0\,.
\label{plmm}
\eeqn

Keeping only the second term gives
\begin{widetext}
\beqn
\delta \left(\partial_tv_l\right)&=&2i\lambda bD\left(\frac{d^2-2}{d(d+2)}\right)P_{lmn}(\bk)\left(\delta_{jn}\delta_{ms}+\delta_{ns}\delta_{mj}\right)v_j(\tbh)v_s(\tbk-\tbh)\int_{\bq,\Omega}\frac{\mu q^2}{[\Omega^2+\mu^2 q^4]^2}\nonumber\\&=&\frac{i\lambda bDS_d\Lambda^{d-4}d\ell}{2\mu^2}\left(\frac{d^2-2}{d(d+2)}\right)P_{lmn}(\bk)\left[v_n(\tbh)v_m(\tbk-\tbh)+v_m(\tbh)v_n(\tbk-\tbh)\right]\nonumber\\&=&\frac{i\lambda bDS_d\Lambda^{d-4}d\ell}{\mu^2}\left(\frac{d^2-2}{d(d+2)}\right)P_{lmn}(\bk)v_m(\tbh)v_n(\tbk-\tbh)\,,
\label{III.3}
\eeqn
\end{widetext}
where in the last step we have used the symmetry of $P_{lmn}(\bk)$ under interchange of its last two indices.

This is immediately recognized as a contribution to the $-{i\over 2}\lambda$ term in Eq. (3) in the main text, which implies a correction to $\lambda$ given by
\beqn
\delta\lambda=\frac{-2\lambda bDS_d\Lambda^{d-4}d\ell}{\mu^2}\left(\frac{d^2-2}{d(d+2)}\right)\,,
\label{dellambda_IV}
\eeqn
which is just the first term on the RHS of Eq. (\ref{dellambdaIII})}.

\subsection{The Second Graph in (c)}
This graph represents an additional
contribution $\delta \left(\partial_tv_l\right)$ to $ \partial_tv_l$ given by
\begin{widetext}
\beqn
\delta \left(\partial_tv_l\right)=12\left(\frac{-\ii\lambda}{2}\right)\left(-\frac{b}{3}\right) 2DQ_{lmno}(\bk)\int_{\tbh} v_j(\tbh)v_o(\tbk-\tbh)
\int_{\bq, \Omega} \frac{P_{mi}(\bq)P_{nij}(\bh-\bq)}{(\mu^2q^4+\Omega^2)[\mu |\bh-\bq|^2-\ii(\omega-\Omega)]}\ .
\label{dvIVstart}
\eeqn
\end{widetext}
where the combinatoric prefactor of $12$ arises because there are $3$ ways to pick the leg with index $o$ on the left. Then, once this choice has been made, there are $2$ ways to pick the leg with index $m$ on the left, and $2$ ways to pick the one with index $i$ on the right.

To extract from (\ref{dvIVstart}) the contribution to the  $-{i\over 2}\lambda$ term in Eq. (3) in the main text, we need the pieces of $\delta\left( \partial_tv_l\right)$ that are linear in either the external momenta $\bk$ or $\bh$. We notice that the integral over $\tbq$ vanishes if we set $\bh=0$ in its integrand. This implies the entire term is at least of order $\bh$. Thus, to obtain a correction to $\lambda$ we can simply set the external frequency $\omega=0$ in the integrand; doing so, and integrating over $\Omega$, we obtain
\begin{widetext}
\beqn
\delta \left(\partial_tv_l\right)=2\ii\lambda b DQ_{lmno}(\bk)\int_{\tbh} v_j(\tbh)v_o(\tbk-\tbh)
\int_{\bq} \frac{P_{mi}(\bq)}{\mu q^2[\mu q^2+\mu |\bh-\bq|^2]}\left[P_{ni}(\bh-\bq)(h_j-q_j)+P_{nj}(\bh-\bq)(h_i-q_i)\right]\,.\nonumber\\
\label{Graphd1}
\eeqn
\end{widetext}
This can be rewritten as
\begin{widetext}
\beqn
\delta\left( \partial_tv_l\right)=2\ii\lambda b DQ_{lmno}(\bk)\int_{\tbh} v_j(\tbh)v_o(\tbk-\tbh)
\left(I^{(2)}_{jmn}(\bh)+I^{(3)}_{jmn}(\bh)+I^{(4)}_{jmn}(\bh)+I^{(5)}_{jmn}(\bh)\right)
\label{dvIV.2}
\eeqn
\end{widetext}
where $I^{(2)}_{jmn}(\bh)$ is given by equation (\ref{I2}) (with the obvious substitution $\bk\to\bh$), and the other integrals are defined as:\begin{eqnarray}
I^{(3)}_{jmn}(\bh)\equiv h_j\int_{\bq} \frac{P_{mi}(\bq)P_{ni}(\bh-\bq)}{\mu q^2[\mu q^2+\mu |\bh-\bq|^2]},\\
I^{(4)}_{jmn}(\bh)\equiv h_i\int_{\bq} \frac{P_{mi}(\bq)P_{nj}(\bh-\bq)}{\mu q^2[\mu q^2+\mu |\bh-\bq|^2]},\\
I^{(5)}_{jmn}(\bh)\equiv -\int_{\bq} \frac{P_{mi}(\bq)P_{nj}(\bh-\bq)q_i}{\mu q^2[\mu q^2+\mu |\bh-\bq|^2]}.
\end{eqnarray}
Immediately, by the properties of the projection operator we get
\beqn
I^{(5)}_{jmn}(\bh)=0.
\label{I5}
\eeqn
We notice  that both $I^{(3)}_{jmn}(\bh)$ and $I^{(4)}_{jmn}(\bh)$ are already proportional to $\bh$, so we can simply set $\bh={\bf 0}$ inside the integral. Thus, we obtain
\begin{widetext}
\begin{eqnarray}
I^{(3)}_{jmn}(\bh)&=& h_j\int_{\bq} \frac{P_{mi}(\bq)P_{ni}(\bq)}{2\mu^2 q^4}= h_j\int_{\bq} \frac{P_{mn}(\bq)}{2\mu^2 q^4}= h_j\int_{\bq} \frac{\langle P_{mn}(\bq)\rangle_{_{\bf\hat{q}}}}{2\mu^2 q^4}
     = {d-1\over 2d}{S_d\over\left(2\pi\right)^d}\mu^{-2}\Lambda^{d-4}d\ell\,h_j\delta_{mn},\label{I3}\\
I^{(4)}_{jmn}(\bh)&=& h_i\int_{\bq} \frac{P_{mi}(\bq)P_{nj}(\bq)}{2\mu^2 q^4}=h_i\int_{\bq}
\frac{\langle P_{mi}(\bq)P_{nj}(\bq)\rangle_{_{\bf\hat{q}}}}{2\mu^2 q^4}
\nonumber\\
&=& {1\over 2d(d+2)}{S_d\over\left(2\pi\right)^d}\mu^{-2}\Lambda^{d-4}d\ell\,\left[\left(d^2-3\right)\delta_{nj}h_m+\delta_{mn}h_j+\delta_{mj}h_n\right].\label{I4}
\end{eqnarray}
\end{widetext}

Plugging the values (\ref{I2}),  (\ref{I3}), (\ref{I4}),  and (\ref{I5}) of the various integrals into Eq. (\ref{dvIV.2}), we obtain
\begin{widetext}
\begin{eqnarray}
\delta \left(\partial_tv_l\right)= \ii{\lambda b D\over 2\mu^2}{S_d\over\left(2\pi\right)^d}\Lambda^{d-4}d\ell Q_{lmno}(\bk)\int_{\tbh} v_j(\tbh)v_o(\tbk-\tbh)\left[{1\over d}\delta_{mj}h_n+{2d^2-d-10\over d(d+2)}\delta_{nj}h_m+{d+1\over d}\delta_{mn}h_j\right].
\end{eqnarray}
\end{widetext}
The third piece vanishes due to the incompressibility condition $h_jv_j(\tbh)=0$. The first and the second pieces can be grouped together since $Q_{lmno}$ is invariant under the interchange of $m$ and $n$. Therefore, the above expression can be simplified as
\begin{widetext}
\begin{eqnarray}
\delta\left( \partial_tv_l\right)=&&\ii{(d-2)\over d}{\lambda b D\over \mu^2}{S_d\over\left(2\pi\right)^d}\Lambda^{d-4}d\ell Q_{lmno}(\bk)\int_{\tbh} v_j(\tbh)v_o(\tbk-\tbh)\delta_{mj}h_n\nonumber\\
=&&\ii {(d-2)\over d}{\lambda b D\over \mu^2}{S_d\over\left(2\pi\right)^d}\Lambda^{d-4}d\ell \int_{\tbh} \left[P_{lj}(\bk)h_o+P_{\ell n}(\bk)h_n\delta_{jo}+P_{\ell o}(\bk)h_j\right]v_j(\tbh)v_o(\tbk-\tbh)\nonumber\\
=&&\ii{(d-2)\over d}{\lambda b D\over \mu^2}{S_d\over\left(2\pi\right)^d}\Lambda^{d-4}d\ell \int_{\tbh} P_{lj}(\bk)h_ov_j(\tbh)v_o(\tbk-\tbh)\nonumber\\
=&&\ii{(d-2)\over d}{\lambda b D\over \mu^2}{S_d\over\left(2\pi\right)^d}\Lambda^{d-4}d\ell \int_{\tbh} P_{lj}(\bk)k_ov_j(\tbh)v_o(\tbk-\tbh)\nonumber\\
=&&\ii{d-2\over 2d}{\lambda b D\over \mu^2}{S_d\over\left(2\pi\right)^d}\Lambda^{d-4}d\ell \int_{\tbh} P_{ljo}(\bk)v_j(\tbh)v_o(\tbk-\tbh),\label{Diagramd}
\end{eqnarray}
\end{widetext}
where in the second equality we have dropped the second and third pieces. The second piece can be dropped because it vanishes, as can be seen by simply changing variables of integration from ${\bf h}$ to ${\bf k}-{\bf h}$; this gives
\beqn
\nonumber
&&\int_{\tbh}P_{\ell n}(\bk)h_nv_o(\tbk-\tbh)v_o(\bh)
\\
\nonumber
&=&\int_{\tbh}P_{\ell n}(\bk)(k_n-h_n)v_o(\tbk-\tbh)v_o(\tbh)\,.
\eeqn
Adding the left and the right hand side of this equation, and dividing by $2$, implies
\begin{widetext}
\begin{eqnarray}
\int_{\tbh}P_{\ell n}(\bk)h_nv_o(\tbk-\tbh)v_o(\tbh)&=&{1\over 2}\left(\int_{\tbh}P_{\ell n}
(\bk)h_nv_o(\tbk-\tbh)v_o(\tbh)+\int_{\tbh}P_{\ell n}(\bk)(k_n-h_n)v_o(\tbk-\tbh)v_o(\tbh)
\right)\nonumber\\
&=&{1\over 2}\int_{\tbh}P_{\ell n}(\bk)k_nv_o(\tbk-\tbh)v_o(\tbh)=0\,\,,
\end{eqnarray}
\end{widetext}
with the last equality following from 
$P_{\ell n}(\bk)k_n=0$.
The third piece can be dropped due to the incompressibility condition $h_jv_j(\tbh)=0$. The remaining piece of  (\ref{Diagramd}) is readily recognized as a contribution to the $-\frac{i}{2}\lambda$ term in Eq. (3) in the main text. This implies a correction to $\lambda$ given by the second piece on the RHS of Eq. (\ref{dellambdaIV}).

\subsection{Graph (d)}

This graph represents an additional
contribution $\delta \left(\partial_tv_l\right)$ to $ \partial_tv_l$ given by:
\begin{widetext}
\beqn
\delta \left(\partial_tv_l\right)&=&-\lambda^2P_{lmn}(\bk)v_j(\bk)\int_{\bq, \Omega} \frac{2DP_{mi}(\bq)P_{nij}(\bk-\bq)}{(\mu^2q^4+\Omega^2)[\mu|\bk-\bq|^2-\ii(\omega-\Omega)]}\nonumber\\
&=&-D\lambda^2P_{lmn}(\bk)v_j(\bk)\int_{\bq} \frac{P_{mi}(\bq)P_{nij}(\bk-\bq)}{\mu q^2[\mu|\bk-\bq|^2-\ii\omega+\mu q^2]}\,.
\label{B2}
\eeqn
\end{widetext}
The integral in this expression is  readily seen to vanish when $\bk\rightarrow {\bf 0}$, since the integrand then becomes odd in $\bq$. Hence, the integral is at least of order $\bk$, so the entire term (include the implicit first power of $\bk$ coming from the $P_{lmn}(\bk)$ in front), is $\cO(k^2)$. Since we do not need to keep any terms in the equations of motion higher order in $\bk$ and $\omega$ than  $\cO(k^2)$, this means that we can safely set $\omega=0$ inside the integral. Keeping just the  $\cO(k)$ piece of the integral then gives us a modification to the equation of motion of $\cO(k^2 \bv)$, which is clearly a renormalization of the diffusion constant $\mu$. So setting $\omega=0$ in the integrand for the reasons just discussed, and then  writing $P_{nij}(\bk-\bq)$ using  its definition as given in the main text after Eq. (3), gives for the integral in (\ref{B2}):
\begin{widetext}
\beqn
\int_{\bq} \frac{P_{im}(\bq)P_{nij}(\bk-\bq)}{\mu q^2[\mu|\bk-\bq|^2-\ii\omega+\mu q^2]}=\int_{\bq}
\frac{P_{mi}(\bq)\left[P_{jn}(\bq-\bk)(k_i-q_i)+P_{in}(\bq-\bk)(k_j-q_j)\right]}{\mu q^2[\mu|\bk-\bq|^2+\mu q^2]}\,.
\label{IIint1}
\eeqn
\end{widetext}
The term proportional to $k_j$ in this expression can be dropped, since $k_jv_j=0$ (this is just the incompressibility condition ${\bf\nabla}\cdot\bv=0$ written in Fourier space). The term proportional
to $q_i$ can be dropped since $P_{mi}(\bq)q_i=0$ by the properties of the transverse projection operator $P_{mi}(\bq)$. This leaves two terms in the integral, which can be written as
\beqn
I^{(1)}_{jmn}(\bk)\equiv k_i\int_{\bq}
\frac{P_{mi}(\bq)P_{jn}(\bq-\bk)}{\mu q^2[\mu|\bk-\bq|^2+\mu q^2]}\,,
\label{I1}
\eeqn
and
\beqn
I^{(2)}_{jmn}(\bk)\equiv -\int_{\bq} \frac{P_{mi}(\bq)P_{ni}(\bk-\bq)q_j}{\mu q^2[\mu q^2+\mu |\bk-\bq|^2]}\,.
\label{I2}
\eeqn
Since $I^{(1)}_{jmn}(\bk)$ already has an explicit factor of $k$ in front, we can evaluate it to
linear order in $k$ by setting $\bk={\bf 0}$ inside the integral. Doing so gives
\beqn
I^{(1)}_{jmn}(\bk)={1\over2\mu^2}k_i\int_{\bq}
\frac{P_{mi}(\bq)P_{jn}(\bq)}{ q^4}\,,
\label{I1simple}
\eeqn
The integral in this expression can now be evaluated by replacing the only piece that depends on the direction ${\hat{\bf q}}$ of $\bq$, namely, the factor  $P_{mi}(\bq)P_{jn}(\bq)$, with its angle average. As shown in  (\ref{Sec: <}), this average is given by
\begin{widetext}
\beqn
\langle P_{mi}(\bq)P_{jn}(\bq)\rangle_{_{\hat{\bf q}}}=\left({d^2-3\over d(d+2)}\right)\delta_{mi}\delta_{jn}+{1\over d(d+2)}(\delta_{mn}\delta_{ij} +\delta_{mj}\delta_{ni})\,.
\label{P2angle}
\eeqn
\end{widetext}
Inserting this into (\ref{I1simple}) gives
\begin{widetext}
\beqn
I^{(1)}_{jmn}(\bk)={1\over2\mu^2}\left[\left({d^2-3\over d(d+2)}\right)k_m\delta_{jn}+{1\over d(d+2)}\left(k_j\delta_{mn}+k_n\delta_{mj}\right)\right]{S_d\over(2\pi)^d}\Lambda^{d-4}d\ell \,.
\label{I1final}
\eeqn
\end{widetext}

Now let us expand $I^{(2)}_{jmn}(\bk)$ to linear order in $k$. 
Changing  variables of integration from $\bq$ to a shifted variable $\bp$ defined by:
\begin{eqnarray}
 \bq=\bp+{\bk\over 2}\,
\end{eqnarray}
gives
\begin{widetext}
\begin{eqnarray}
  I^{(2)}_{jmn}(\bk)&= &-\int_{\bp} \frac{P_{mi}(\bp_+)P_{ni}(\bp_-)p_j}{\Gamma(\bp_+)[\Gamma(\bp_+)+\Gamma(\bp_-)]}-{k_j\over 2}\int_{\bp} \frac{P_{mi}(\bp_+)P_{ni}(\bp_-)}{\Gamma(\bp_+)[\Gamma(\bp_+)+\Gamma(\bp_-)]}\nonumber\\
  &\equiv& I^{(2.1)}_{jmn}(\bk)+ I^{(2.2)}_{jmn}(\bk),
  \label{I2ab}
\end{eqnarray}
\end{widetext}
where we've defined
\begin{eqnarray}
 \bp_+\equiv\bp+{\bk\over 2},\,\,\,\,\,\,\,\,\bp_-\equiv\bp-{\bk\over 2},
\end{eqnarray}
and $I^{(2.1)}_{jmn}(\bk)$ and $I^{(2.2)}_{jmn}(\bk)$ to be the first and second terms in the expression for $I^{(2)}_{jmn}(\bk)$.
Since $I^{(2.2)}_{jmn}(\bk)$ has an explicit factor of $k$ in front, we can evaluate this term to linear order in $\bk$ by setting $\bk=0$ inside the integral. This leads to
\begin{widetext}
\begin{eqnarray}
I^{(2.2)}_{jmn}(\bk)=-{k_j\over 2}\int_{\bp} \frac{P_{mi}(\bp)P_{ni}(\bp)}{\Gamma(\bp)[\Gamma(\bp)+\Gamma(\bp)]}&=&-{d-1\over 4d}{S_d\over\left(2\pi\right)^d}\mu^{-2}\Lambda^{d-4}d\ell\,k_j\delta_{mn}.\label{I2b}
\end{eqnarray}
\end{widetext}
The calculation of $I^{(2.1)}_{jmn}(\bk)$ requires more effort. Expanding the numerator we get
\begin{widetext}
\begin{eqnarray}
   I^{(2.1)}_{jmn}(\bk)&=&-\int_{\bp} \frac{p_j}{\Gamma(\bp_+)}\frac{1}{[\Gamma(\bp_+)+\Gamma(\bp_-)]}\left(\delta_{mi}-{p^+_mp^+_i\over p_+^2}\right)\left(\delta_{ni}-{p^-_np^-_i\over p_-^2}\right)\nonumber\\
      &=& -\int_{\bp} \frac{p_j}{\Gamma(\bp_+)}\frac{\delta_{mn}-{p^-_mp^-_n\over p_-^2}-{p^+_mp^+_n\over p_+^2}}{[\Gamma(\bp_+)+\Gamma(\bp_-)]}-\int_{\bp} \frac{p_jp_m^+p_n^-}{\Gamma(\bp_+)}\frac{\bp_+\cdot\bp_-}{[\Gamma(\bp_+)+\Gamma(\bp_-)]p_-^2p_+^2}.\label{I2a1}
\end{eqnarray}
\end{widetext}
Note that we have purposely written each term on the RHS of the second equality in Eq. (\ref{I2a1}) as an even function of $\bk$ multiplied by a non-even function. We can simply set $\bk=0$ inside the even part since it cannot be expanded to given a linear piece in $k$. Therefore, Eq. (\ref{I2a1}) can be simplified as
\begin{widetext}
\begin{eqnarray}
 I^{(2.1)}_{jmn}(\bk)&=&-\int_{\bp} \frac{p_j}{\Gamma(\bp_+)}\frac{p^2\delta_{mn}-2p_mp_n}{2\Gamma(\bp)p^2}-\int_{\bp} \frac{p_jp_m^+p_n^-}{\Gamma(\bp_+)}\frac{1}{2\Gamma(\bp)p^2}.\nonumber\\
      &=&-\int_{\bp} \frac{p_j}{\Gamma(\bp_+)}\frac{p^2\delta_{mn}-2p_mp_n}{2\Gamma(\bp)p^2}-\int_{\bp} \frac{p_j}{\Gamma(\bp_+)}\frac{1}{2\Gamma(\bp)p^2}\left(p_mp_n-{p_mk_n\over 2}+{p_nk_m\over 2}-{k_mk_n\over 4}\right)\nonumber\\
      &=&-\int_{\bp} \frac{p_j}{\Gamma(\bp_+)}\frac{1}{2\Gamma(\bp)p^2}\left(p^2\delta_{mn}-p_mp_n-{p_mk_n\over 2}+{p_nk_m\over 2}-{k_mk_n\over 4}\right)\nonumber\\
     &=&-{1\over 2\mu^2}\int_{\bp} \frac{p_j}{p^6}\left(1-{p_sk_s\over p^2}\right)\left(p^2\delta_{mn}-p_mp_n-{p_mk_n\over 2}+{p_nk_m\over 2}\right)+O(k^2)\nonumber
     \\
    &=& -{1\over 2\mu^2}\int_{\bp} \frac{1}{p^6}\left( -\frac{p_jp_mk_n}{2}+\frac{p_jp_nk_m}{2}-\delta_{mn}p_jp_sk_s + \frac{p_jp_sp_mp_nk_s}{p^2}
    \right)+O(k^2)\,.
\label{I2a2.1}
\end{eqnarray}
\end{widetext}

The integral over $\bp$ in this expression can now be evaluated by replacing $p_jp_m$, $p_jp_n$,  $p_jp_s$, and  $p_jp_sp_mp_n$ with their angular averages over all directions of $\bp$ for fixed $|\bp|$, as given by
equations (\ref{angleq2}) and (\ref{q4angle2}) of section (\ref{Sec: <}). This gives
\begin{widetext}
\begin{eqnarray}
    I^{(2.1)}_{jmn}(\bk)&=& -{1\over 2\mu^2}{S_d\over\left(2\pi\right)^d}\Lambda^{d-4}d\ell \left( -\frac{\delta_{mj}k_n}{2d}+\frac{\delta_{jn}k_m}{2d}-\frac{\delta_{mn}k_j}{d} + \frac{\delta_{mn}k_j+\delta_{nj}k_m+\delta_{mj}k_n}{d(d+2)}
    \right)+O(k^2)\nonumber
    \\
 &=& -{1\over 2\mu^2}{S_d\over\left(2\pi\right)^d}\Lambda^{d-4}d\ell \left(-\frac{1}{2(d+2)}\delta_{mj}k_n -\frac{d+1}{d(d+2)}\delta_{mn}k_j+\frac{d+4}{2d(d+2)}\delta_{nj}k_m
    \right)+O(k^2)\nonumber
    \\
    &=&  {1\over4d(d+2)}{S_d\over\left(2\pi\right)^d}\mu^{-2}\Lambda^{d-4}d\ell\left[d\delta_{mj}k_n-(d+4)\delta_{nj}k_m+2(d+1)\delta_{mn}k_j\right]+O(k^2).
\label{I2a2}
\end{eqnarray}
\end{widetext}
Plugging Eqs. (\ref{I2b},\ref{I2a2}) into Eq. (\ref{I2ab}) we get
\begin{widetext}
\begin{eqnarray}
 I^{(2)}_{jmn}(\bk)={1\over4d(d+2)}{S_d\over\left(2\pi\right)^d}\mu^{-2}\Lambda^{d-4}d\ell\left[d\delta_{mj}k_n-(d+4)\delta_{nj}k_m+(-d^2+d+4)\delta_{mn}k_j\right].
\label{I2}
\end{eqnarray}
\end{widetext}

The terms in $I^{(1)}_{jmn}(\bk)$ (\ref{I1final}) and $I^{(2)}_{jmn}(\bk)$ (\ref{I2}) that are proportional to $k_j$ may be dropped, since they multiply $v_j(\tbk)$, and, hence, vanish by the incompressibility condition $k_jv_j=0$. Dropping them, and
adding these two integrals $I^{(1)}_{jmn}(\bk)$ and $I^{(2)}_{jmn}(\bk)$ makes the entire correction to the equation of motion coming from graph $II$ become:
\begin{widetext}
\beqn
\delta \left(\partial_tv_l\right)&=&-D\lambda^2
P_{lmn}(\bk)v_j(\tbk)\left(I^{(1)}_{jmn}(\bk)+I^{(2)}_{jmn}(\bk)\right)=-{D\lambda^2\over 2\mu^2}{S_d\over (2\pi)^d}\Lambda^{d-4} d\ell
P_{lmn}(\bk)v_j(\tbk)\nonumber\\&\times&\left[\left(1-\frac{2}{d}+\frac{1}{d(d+2)}-\frac{d+4}{2d(d+2)}\right)k_m\delta_{jn}+\left(\frac{1}{d(d+2)}+\frac{d}{2d(d+2)}\right)k_n\delta_{jm}\right]\,.
\label{dvII.1}
\eeqn
\end{widetext}
Simplifying, and performing the tensor index contractions, gives
\begin{widetext}
\beqn
\delta\left( \partial_tv_l\right)&=&-{D\lambda^2\over 2\mu^2}{S_d\over (2\pi)^d}\Lambda^{d-4} d\ell \left[
P_{lmj}(\bk)k_m\left(1-\frac{5}{2d}\right)+P_{ljn}(\bk)k_n\frac{1}{2d}\right]v_j(\tbk)\nonumber\\&=&-{D\lambda^2\over 2\mu^2}{S_d\over (2\pi)^d}\Lambda^{d-4} d\ell \left[
P_{lmj}(\bk)k_m\left(1-\frac{2}{d}\right)\right]v_j(\tbk)\,,
\label{dvII.2}
\eeqn
\end{widetext}
where in the second step we have used the symmetry of $P_{ljn}(\bk)$ to write $P_{ljn}(\bk)k_n=
P_{lmj}(\bk)k_m$. Now from the definition of $P_{lmj}(\bk)$, we have $P_{lmj}(\bk)k_m=P_{lm}(\bk)k_jk_m+P_{lj}(\bk)k_mk_m$. The first term in this expression vanishes by the fundamental property of the transverse projection operator, while the second is just $k^2 P_{lj}(\bk)$. Thus, we finally obtain
\begin{widetext}
\beqn
\delta \left(\partial_tv_l\right)&=&-\left(\left(\frac{d-2}{d}\right){D\lambda^2\over 2\mu^2}{S_d\over (2\pi)^d}\Lambda^{d-4} d\ell\right) k^2 P_{lj}(\bk)v_j(\tbk)\,,
\label{dvIIfinal}
\eeqn
\end{widetext}
which is exactly what one would get by adding to $\mu$ (hiding in the ``propagator'') in Eq. (3) in the main text 
 a correction $\delta\mu$ given by Eq. (\ref{delmu}).

\subsection{Vanishing diagrams}

Besides the five non-vanishing one-loop diagrams, there are also two sets of one-loop diagrams that cancel exactly, giving zero contribution to the corrections (Fig.~\ref{fig:pic}).

\begin{figure}
\centering
\begin{tabular}{c}
\includegraphics[width=0.45\textwidth]{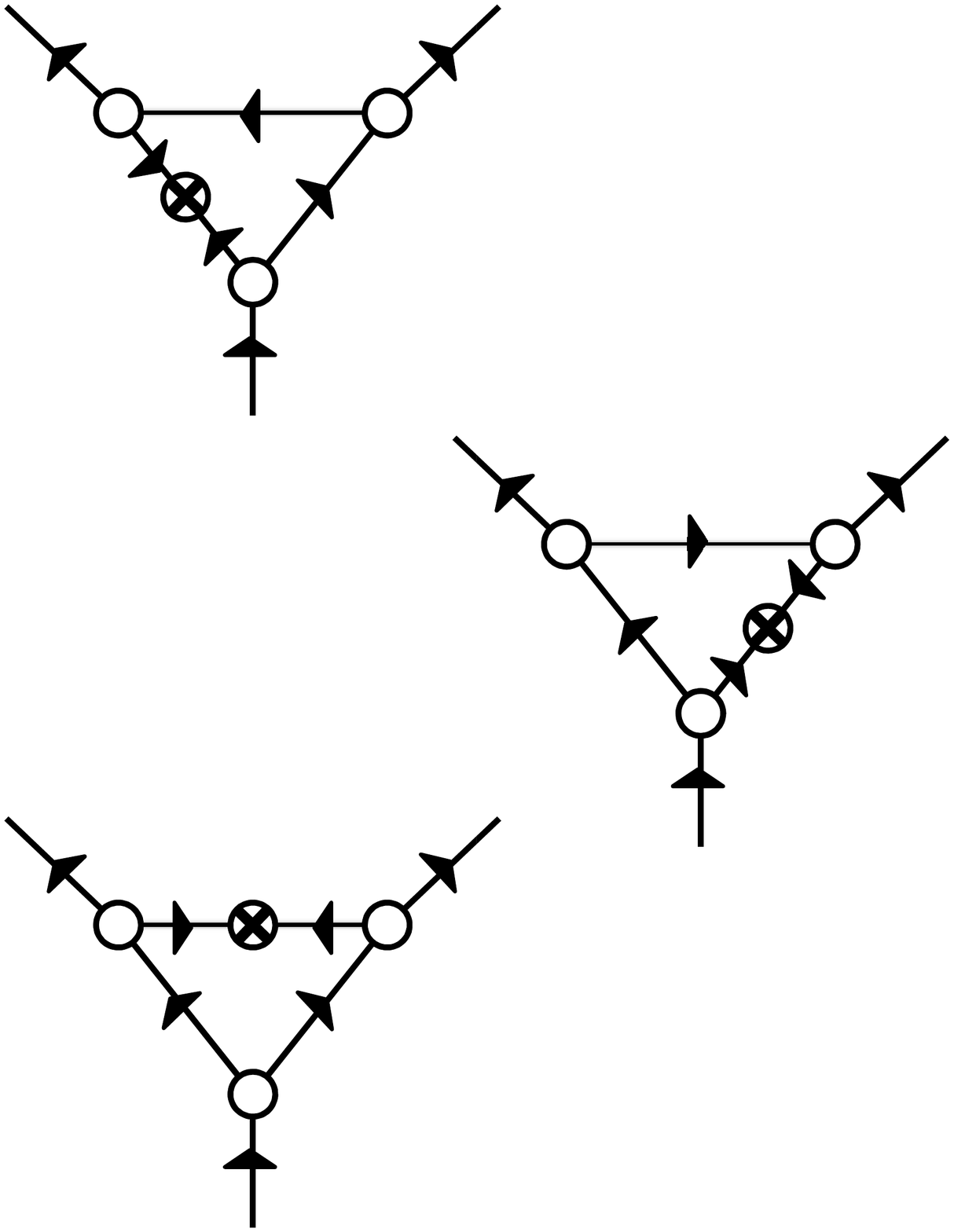} \\
\hline
\includegraphics[width=0.45\textwidth]{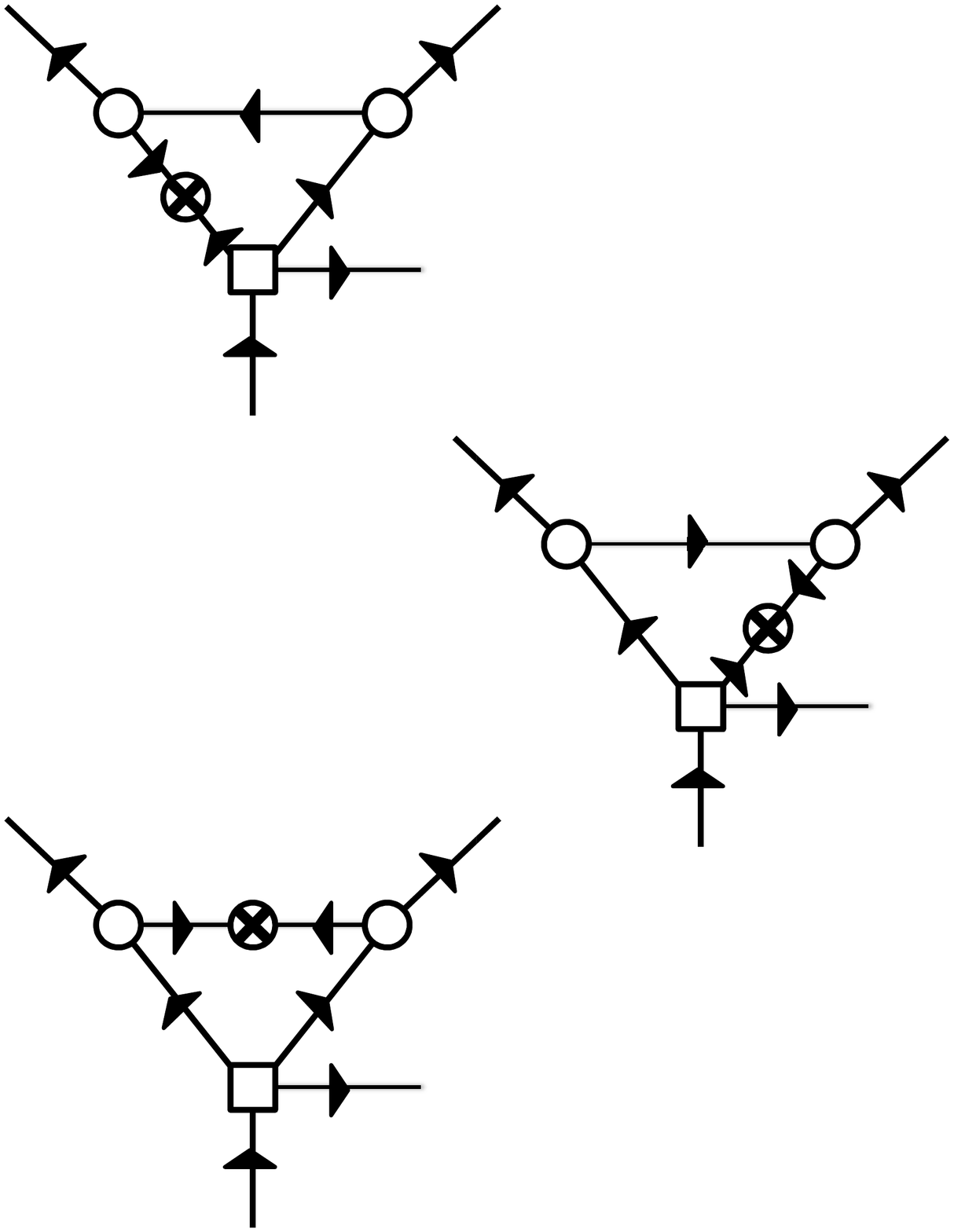} 
\end{tabular}
\caption{ Vanishing sets of diagrams. {\it Top}: The three diagrams consist of three three-point vertices. They cancel each other, leading to zero net contribution to $\lambda$. {\it Bottom}: The three diagrams consist of one four-point vertex and two three-point vertices. They again cancel each other, leading to zero net contribution to $b$.}
\label{fig:pic}
\end{figure}

%
%

\section{\label{Sec: <}Angular Averages}
In this section we derive the various angular averages used in the previous sections.

We begin by deriving the identity
\beqn
\langle {q_mq_n\over q^2}\rangle_{_{\hat{\bf q}}}={1\over d}\delta_{mn}\,,
\label{angleq2}
\eeqn
where $\langle \rangle_{_{\hat{\bf q}}}$
denotes the average over directions ${\hat{\bf q}}$ of $\bq$ for fixed $|\bq|$. The identity (\ref{angleq2}) follows by symmetry: the average in question clearly must vanish when $m\ne n$, since then the quantity being averaged is odd in $\bq$. Furthermore, when $m=n$, the average must be independent of the value that $m$ and $n$ both equal. Hence, this average must be proportional to
$\delta_{mn}$.

The constant of proportionality in (\ref{angleq2}) is easily determined by noting that the trace of this average over $mn$ is $\langle {q_mq_m\over q^2}\rangle_{_{\hat{\bf q}}}=\langle{q^2\over q^2}\rangle_{_{\hat{\bf q}}}=\langle {1}\rangle_{_{\hat{\bf q}}}=1$. This forces the prefactor of ${1\over d}$ in (\ref{angleq2}).

From (\ref{angleq2}), it obviously follows that
\beqn
\langle P_{mn}\rangle_{_{\hat{\bf q}}}=\langle {\delta_{mn}-{q_mq_n\over q^2}}\rangle_{_{\hat{\bf q}}}=(1-{1\over d})\delta_{mn}\,,
\label{P1}
\eeqn
which is the identity we used in equation (\ref{dela}).

We now consider the average of {\it two} projection operators
$\langle P_{mi}(\bq)P_{jn}(\bq)\rangle_{_{\hat{\bf q}}}$ that appears in (\ref{P2angle}). Using the definition of the projection operator, this can be written as follows:
\begin{widetext}
\beqn
\left< P_{mi}(\bq)P_{jn}(\bq)\right>_{_{\hat{\bf q}}}=\langle \left(\delta_{mi}-{q_mq_i\over q^2}\right)\left(\delta_{jn}-{q_jq_n\over q^2}\right)\rangle_{_{\hat{\bf q}}}=\delta_{mi}\delta_{jn}-\delta_{mi}\langle{q_jq_n\over q^2}\rangle_{_{\hat{\bf q}}}-\delta_{jn}\langle{q_mq_i\over q^2}\rangle_{_{\hat{\bf q}}}+\langle{q_iq_jq_mq_n\over q^4}\rangle_{_{\hat{\bf q}}}\,.
\label{P2angleexp}
\eeqn
\end{widetext}
The first two angular averages on the right hand side of this expression can be read off from (\ref{angleq2}). The last is new, and can be evaluated as follows:

First, note that
by symmetry, this average vanishes unless the four indices $i$, $j$, $k$, $l$, are equal in pairs. Furthermore, if they are equal in pairs, but the pairs are different (e.g., if $i=j=x$ and $m=n=z$), then the average will have one value, independent of what the values of the two pairs of indices are (e.g., if $i=j=y$ and $m=n=x$, the average would be the same as in the example just cited.
The only other non-zero possibility is that {\it all four} indices are equal, in which case the average is the same no matter which index all four are equal to (i.e., the average when $i=j=m=n=x$ is the same as that when  $i=j=m=n=z$). Furthermore, this average must be completely symmetric under {\it any} interchange of its indices. This can be summarized by saying that the average must take the form:
\beqn
&&\langle{q_iq_jq_mq_n\over q^4}\rangle_{_{\hat{\bf q}}}\nonumber\\
&=&A\Upsilon_{ijmn}+B(\delta_{mi}\delta_{jn}+\delta_{ij}\delta_{mn}+\delta_{in}\delta_{jm})\,,
\label{q4angle}
\eeqn
where $\Upsilon_{ijmn}=1$ if and only if $i=j=m=n$, and is zero otherwise, and $A$ and $B$ are unknown, dimension ($d$) dependent constants that we'll now determine.

We
can derive one condition on $A$ and $B$
by taking the trace of (\ref{q4angle}) over any two indices (say, $i$ and $j$). This gives
\beqn
\langle {q_mq_n\over q^2}\rangle_{_{\hat{\bf q}}}=(A+(d+2)B)\delta_{mn}\,.
\eeqn
Comparing this with (\ref{angleq2}) gives
\beqn
A+(d+2)B={1\over d}\,.
\label{ABcond1}
\eeqn
A second condition can be derived by explicitly evaluating the angle average when all four indices are equal. Since it doesn't matter what value they all equal, we'll chose it to be $z$. Defining $\theta$ to be the angle between the $z$-axis and $\bq$,  we can obtain the needed average in $d$-dimensions by integrating in hyperspherical coordinates:
\beqn
\langle {q_z^4\over q^4}\rangle_{_{\hat{\bf q}}}={\int_0^\pi d\theta
\cos^4\theta \sin^{d-2}\theta\over\int_0^\pi d\theta \theta \sin^{d-2}\theta}={3\over d(d+2)}\,.
\eeqn
Comparing this with (\ref{q4angle}) evaluated for $i=j=m=m$  gives
\beqn
A+3B={3\over d(d+2)}\,.
\label{ABcond2}
\eeqn
The simultaneous solution for $A$ and $B$ of equations (\ref{ABcond1}) and (\ref{ABcond2})
is $A=0$ and $B={1\over d(d+2)}$. Using these  in (\ref{q4angle}) gives
\beqn
\langle{q_iq_jq_mq_n\over q^4}\rangle_{_{\hat{\bf q}}}={1\over d(d+2)}(\delta_{mi}\delta_{jn}+\delta_{ij}\delta_{mn}+\delta_{in}\delta_{jm})\,.
\nonumber\\
\label{q4angle2}
\eeqn
Using this and (\ref{angleq2}) in (\ref{P2angleexp}) gives (\ref{P2angle}).

\begin{widetext}
\begin{table}
    \begin{tabular}{ | c | c | c | c|}
    \hline
    Exponents & Incompressible active matter & Heisenberg model \cite{Aharony} & Heisenberg model with dipolar interactions \cite{Zinn} \\  \hline 
    $\beta$ & $0.43 \pm 0.03$ & $0.3645 \pm 0.0025$ & $0.38 \pm 0.02$\\
     $\nu$ & $0.67 \pm 0.04$ & $0.705 \pm 0.003$ & $0.69 \pm 0.02$\\
      $\eta$ & $0.35 \pm 0.08$ & $0.033 \pm 0.004$ & $0.023 \pm 0.015$\\
     $\delta$ & $3.48 \pm 0.03$ & $4.803 \pm 0.037$ & $4.45 \pm 0.04$\\
      $\gamma$ & $1.11 \pm 0.01$ & $1.386 \pm 0.004$ & $1.37 \pm 0.02$\\
    \hline 
        \end{tabular}    
        \caption[Table caption text]{
  Comparisons between the critical exponents obtained in this work and other models in spatial dimension $d=3$.}  
        \label{tab1}
\end{table}
\end{widetext}

\section{Numerical estimates of the critical exponents}

We can estimate the numerical values of the exponents in spatial dimension $d=3$ as follows:
We first choose a scaling relation satisfied by any three exponents (e.g., Eq. (23) in the main text
for $\eta$, $\beta$, and $\nu$). We then determine numerical values for two of them (e.g., $\nu$ and $\beta$) by simply setting $\epsilon=1$ in the $\epsilon$-expansion for them, and dropping the unknown $\cO(\epsilon^2)$ terms.  We now get the value of the third exponent (e.g., $\eta$) by requiring that the scaling law (i.e., Eq. (23) in the main text
) hold exactly in $d=3$. In this example, this gives
$\beta=1/2-6/113\approx.447$,
$\nu=1/2+29/226\approx.628$,
and
$\eta=2\beta/\nu-1\approx.424$.
Next, we take $\eta$ and $\nu$ to be given by their respective $\epsilon$-expansions with $\epsilon= 1$, and get $\beta$ from the exact scaling relation. This gives: $\nu=1/2+29/226\approx.628$,
$\eta=31/113\approx.274$, and $\beta=\nu(1+\eta)/2\approx.400$.
Finally, we take $\beta$ and $\eta$ from their $\epsilon$-expansions, and get $\nu$ from the exact scaling relation,  obtaining $\eta=31/113\approx.274$,
$\beta=1/2-6/113\approx.447$, and
$\nu=2\beta/(1+\eta)=.702$.

Note that each exponent gets two possible values in this approach: one from directly setting $\epsilon=1$ in the $\epsilon$-expansion, and another by obtaining the exponent from the exact scaling relation in $d=3$.

Applying the same approach to Widom's exact scaling relation (i.e., Eq. (30) in the main text)
and the three associated exponents $\gamma$, $\beta$, and $\delta$ gives the possible values: $\beta\approx.447$ or $\beta\approx.457$, $\delta\approx3.451$ or $\delta\approx3.503$, and $\gamma\approx1.119$ or $\gamma\approx1.096$.

So if we look at the range of values we've found for each of the exponents, we have
$.274\le\eta\le.424$,
$.628\le\nu\le.702$,
$.400\le\beta\le.457$, $3.451\le\delta\le3.503$, and $1.096\le\gamma\le1.119$. Assuming, as seems reasonable (and as is true for, e.g., the critical exponents for the equilibrium $\cO(n)$ model \cite{Lubensky}), that the correct values lie within the range spanned by the different approaches we've used here,  we can conclude that, in spatial dimension $d=3$, the critical exponents are as shown in the second column of Table \ref{tab1}.

%
%

%

%
%
%
%

  Comparing the critical exponents with the known values  for the two equilibrium analogs of our system: the  three-dimensional, three component Heisenberg model (i.e., the $\cO(3)$ model) with and without dipolar interactions (third  and fourth columns respectively in Table \ref{tab1}), we see that  $\nu$ and $\beta$ are very close in all three models.  The situation is a little better for $\gamma$ and $\delta$. The biggest difference, however, is clearly in $\eta$, which is much larger in the incompressible flock. Thus experiments to determine this exponent, which, as can be seen from equation (17) in the main text,
can be deduced from velocity correlations right at the critical point, will provide the clearest and most dramatic evidence for the non-equilibrium nature of this system, and the novelty of its universality class.

The values of the exponents in $d=2$ obviously can {\it not} be reliably estimated quantitatively from the $4-\epsilon$-expansion. 
We do note, however, that  the ordered state is expected to exist and to have true long-ranged order. 
This is clear since true long-ranged order exists even in the {\it compressible} problem, which obviously has more fluctuations than the incompressible problem we've studied here. Hence, we do {\it not} expect this problem to be like the {\it equilibrium} $2d$ XY model, in which\cite{KT} the ordered state only has quasi-long-ranged order (i.e., algebraically decaying correlations). We therefore do not expect 2d incompressible flocks to exhibit any of the singular behavior of exponents found in the 2d equilibrium XY model; in particular, there is no reason to expect $\nu=\infty$.
Beyond this, there is little we can say quantitatively about $d=2$ beyond the expectation that the critical exponents $\beta$, $\nu$, $\eta$, $\delta$, and $\gamma$
should be further from their mean field values $\beta=\nu=1/2$, $\eta=0$, $\delta=3$, and $\gamma=1$ than they are in $d=3$. This implies that in $d=2$, $\beta$ will be smaller, and the four other exponents will be bigger, than the values quoted above for $d=3$. 
We also note that the exact scaling relations Eqs.~(23,30) in the main text
will hold in $d=2$, and that all of the exponents will be universal (i.e., the same for all incompressible flocks) in $d=2$.


%


\begin{thebibliography}{2}
\bibitem{comm} This equation of motion is simply what one would obtain from the equation of motion
of references \cite{TT1,TT2,TT3,TT4} by expanding for small $\bv$ and dropping all density-dependent terms.
\bibitem{comm2} The connection between our problem and the dipolar magnet is that the long-ranged dipolar interaction in magnetic systems couples to, and therefore suppresses, the longitudinal component of the magnetization. See \cite{Aharony} for more details.
\end{thebibliography}
\end{document}